\documentclass[preprint,review, 12pt, authoryear]{elsarticle}

\usepackage{amssymb}
\usepackage[hyphens]{url}
\usepackage[breaklinks]{hyperref}   %This is the key package which allows url wrapping.
\usepackage[hyphenbreaks]{breakurl}
\usepackage{longtable}
\usepackage{amsmath,amssymb,multirow,dcolumn,fancyhdr,charter,graphicx}
\usepackage{graphics}
\usepackage{xspace}
\usepackage{color,ulem,epstopdf}
\usepackage[extd]{journalnames} %This is A. Jackson's package that parses ADS journal commands
\usepackage{setspace}
\newcommand{\LL}{\mathcal{L}}

\begin{document}

\begin{frontmatter}

%% Title, authors and addresses

%% use the tnoteref command within \title for footnotes;
%% use the tnotetext command for theassociated footnote;
%% use the fnref command within \author or \address for footnotes;
%% use the fntext command for theassociated footnote;
%% use the corref command within \author for corresponding author footnotes;
%% use the cortext command for theassociated footnote;
%% use the ead command for the email address,
%% and the form \ead[url] for the home page:
%% \title{Title\tnoteref{label1}}
%% \tnotetext[label1]{}
%% \author{Name\corref{cor1}\fnref{label2}}
%% \ead{email address}
%% \ead[url]{home page}
%% \fntext[label2]{}
%% \cortext[cor1]{}
%% \address{Address\fnref{label3}}
%% \fntext[label3]{}

\title{Lunar Crater Identification via Deep Learning}

%% use optional labels to link authors explicitly to addresses:
%% \author[label1,label2]{}
%% \address[label1]{}
%% \address[label2]{}

\author[as,as2,as3,*] {Ari Silburt}
 \author[as,as4,*]{Mohamad Ali-Dib}  \author[as2,as4]{Chenchong Zhu} \author[as,as2,as5]{Alan Jackson} \author[as,as2]{Diana Valencia} \author[as2]{Yevgeni Kissin} \author[as,as4]{Daniel Tamayo} \author[as,as2]{Kristen Menou}

\address[as]{Centre for Planetary Sciences,
    Department of Physical \& Environmental Sciences, University of Toronto Scarborough, Toronto, Ontario M1C 1A4, Canada}
\address[as2]{Department of Astronomy \& Astrophysics,
    University of Toronto, Toronto, Ontario M5S 3H4, Canada}
\address[as3]{Department of Astronomy \& Astrophysics,
    Penn State University, Eberly College of Science, State College, PA 16801, USA}
\address[as4]{Canadian Institute for Theoretical Astrophysics, 60 St. George St, University of Toronto, Toronto, ON M5S 3H8, Canada}
\address[as5]{School of Earth and Space Exploration, Arizona State University, 781 E Terrace Mall, Tempe, AZ 85287-6004, USA}
\address[*]{These authors contributed equally to this work.}

\end{frontmatter}
Keywords: Moon; Crater Detection; Automation; Deep Learning

\noindent Corresponding authors: 
\begin{itemize}
    \item Ari Silburt, Department of Astronomy \& Astrophysics, Penn State University, Eberly College of Science, State College, PA 16801, USA. e-mail: {arisilburt@gmail.com} .
    \item  Mohamad Ali-Dib, Centre for Planetary Sciences, Department of Physical \& Environmental Sciences, University of Toronto Scarborough, Toronto, Ontario M1C 1A4, Canada. e-mail: {m.alidib@utoronto.ca} .
\end{itemize}

\newpage
\section*{Abstract}
Crater counting on the Moon and other bodies is crucial to constrain the dynamical history of the Solar System. This has traditionally been done by visual inspection of images, thus limiting the scope, efficiency, and/or accuracy of retrieval. 
In this paper we demonstrate the viability of using convolutional neural networks (CNNs) to determine the positions and sizes of craters from Lunar digital elevation maps (DEMs). We recover $92\%$ of craters from the human-generated test set and almost double the total number of crater detections. 
Of these new craters, $15\%$ are smaller in diameter than the minimum crater size in the ground-truth dataset. 
Our median fractional longitude, latitude and radius errors are $11\%$ or less, representing good agreement with the human-generated datasets. 
From a manual inspection of 361 new craters we estimate the false positive rate of new craters to be $11\%$. 
Moreover, our Moon-trained CNN performs well when tested on DEM images of Mercury, detecting a large fraction of craters in each map. Our results suggest that deep learning will be a useful tool for rapidly and automatically extracting craters on various Solar System bodies.
We make our code and data publicly available at \url{https://github.com/silburt/DeepMoon.git} and \url{https://doi.org/10.5281/zenodo.1133969}.

%\end{abstract}

%\begin{keyword}
%% keywords here, in the form: keyword \sep keyword

%% PACS codes here, in the form: \PACS code \sep code

%% MSC codes here, in the form: \MSC code \sep code
%% or \MSC[2008] code \sep code (2000 is the default)

%\end{keyword}

%\begin{linenumbers}
%% main text
\section{Introduction}

Craters formed by small impactors constitute an important surface property for many bodies in the Solar System.
On airless bodies like the Moon, Mercury, Ceres, and Vesta, weather based erosion, tectonics and volcanic activity have been largely non-existent resulting in the accumulation of impact craters over time. However, other eroding factors such as micrometeorite bombardment can affect smaller craters. 

Crater densities  permit the geological history of a body to be examined, and the relative chronology of a region to be assessed remotely. 
In addition, when in-situ samples are recovered from a body, absolute chronologies can be determined too. Inferred temporal variation in cratering rates have been used to make inferences about the dynamical history of the Solar System, including the (debated) possibility of a Late Heavy Bombardment, (e.g., \citealt{hartmann1970, ryder2002, gomes2005, malho1,malho2}). Crater records and chronology are thus central to any formation theory about the Solar System. In addition, the size distribution of craters directly probes the dynamics and size distribution of the impactor population \citep{strom2005}. 
For example from the size distribution of craters on the Lunar highlands, \citet{minton2015} argued that the impactor population contained comparatively fewer large bodies than the asteroid belt does today.

Traditionally, crater detection has been done manually via visual inspection of images. 
However this approach is not practical for the vast numbers of kilometre and sub-kilometre sized craters on the Moon (and other Solar System bodies), resulting in human-generated databases that are either spatially comprehensive but restricted to the largest craters, or size comprehensive but limited to a very specific geographic region \citep{stepinski2012, bandeira2012}.
In addition, manual crater counting by experts can yield disagreements as high as 40\% \citep{greeley1970, kirchoff2011, robbins2014}.

As a result, scientists have developed crater detection algorithms (CDAs) to automate the process of classifying craters.  
Such CDAs include edge detection \citep{emami2015}, Hough transforms \citep{salamuniccar2010}, support-vector machines \citep{wetzler2005}, decision trees \citep{stepinski2012} and neural networks \citep{wetzler2005, cohen2016, palafox2017}. 
Multi-step approaches have also been tried. 
For example, \citet{di2014} used a boosting algorithm to box the crater-containing region, then a Hough transform to delimit the edges of the crater. 
\citet{boukercha2014} used a similar approach where an initial detection algorithm provided crater candidates which were subsequently classified as true or false positives by a support-vector machine or polynomial classifier.

These CDAs tend to perform well on the datasets upon which they were trained, but not to generalize well on unseen patches or other bodies (see \citet{stepinski2012} for a review and \citet{chung2014} for a comparison between classical and machine learning based techniques.). 
The difficulty in designing robust CDAs stems from the complex nature of craters, having large variations in shape and illumination, orders of magnitude size differences, overlap and degradation.
An algorithm capable of universally identifying craters on Solar System bodies would be invaluable to the community. 

In this work {we train a deep learning architecture known as a convolutional neural network (CNN)} to perform crater identification\footnote{By crater identification, we mean a) output the pixel locations of crater rims from DEM images via a CNN segmentation task, and b) extract the crater coordinates from these CNN outputs using a custom pipeline, explained in the Methods section.} on Lunar digital elevation map (DEM) images, and transfer-learn our Moon-trained CNN to identify craters on Mercury. 
There are numerous reasons for using CNNs to detect craters.
First, CNNs have demonstrated impressive performance on a variety of computer vision problems and other datasets where features are correlated \citep[e.g.][]{long2015}, demonstrating their versatility.
This includes, in addition to images, sounds and signals. 
Second, CNNs engineer their own representation features, alleviating the need for a human to develop sophisticated pre-processing algorithms and custom input features. 
Finally, CNNs have been able to successfully classify objects that appear at multiple scales in a single image \citep{zhang2016, zeng2017}, a property very relevant to crater counting. 

\section{Methods}
The code to generate the data set (Section~\ref{sec:data}), train our model (Section~\ref{sec:training}), and extract the resulting crater distribution (Section~\ref{sec:craterdet} and Section~\ref{sec:postproc}) is available at \url{https://github.com/silburt/DeepMoon.git}. 
The data used to train, validate and test our model, the global DEM used to make our input images, our best model and final test set crater distribution can be found at \url{https://doi.org/10.5281/zenodo.1133969}.
We use Keras \citep{chollet2015} version 1.2.2 with Tensorflow \citep{abadi2016a} version 0.10 to build and train our model, but our code is also compatible with Keras 2.0.2 and Tensorflow 1.0.

\subsection{Data Preparation}
\label{sec:data}

Our input data was generated by randomly cropping digital elevation map (DEM) images from the Lunar Reconnaisance Orbiter (LRO) and Kaguya merged digital elevation model, which spans $\pm60\, \rm{degrees}$ in latitude (and the full range in longitude) and has a resolution of $512\, \rm{pixels/degree}$, or $59\,\mathrm{m/pixel}$ (\citealt{barker2016}; available at \citealt{lolakaguya2015}). This global grayscale map is a Plate Carree projection with a resolution of $184320 \times 61440\,\mathrm{pixels}$ and a bit depth of $16\,\mathrm{bits/pixel}$; we downsampled it to $92160 \times 30720\,\mathrm{pixels}$ and $8\,\mathrm{bits/pixel}$.  We use an elevation map, rather than an optical one, because a crater's appearance in an elevation map is not affected by the direction of incident sunlight.  This reduces variation in appearance between craters, making it easier to train a CNN to identify them. 

\begin{figure}
\begin{center}
\centerline{\includegraphics[width=1.3\textwidth]{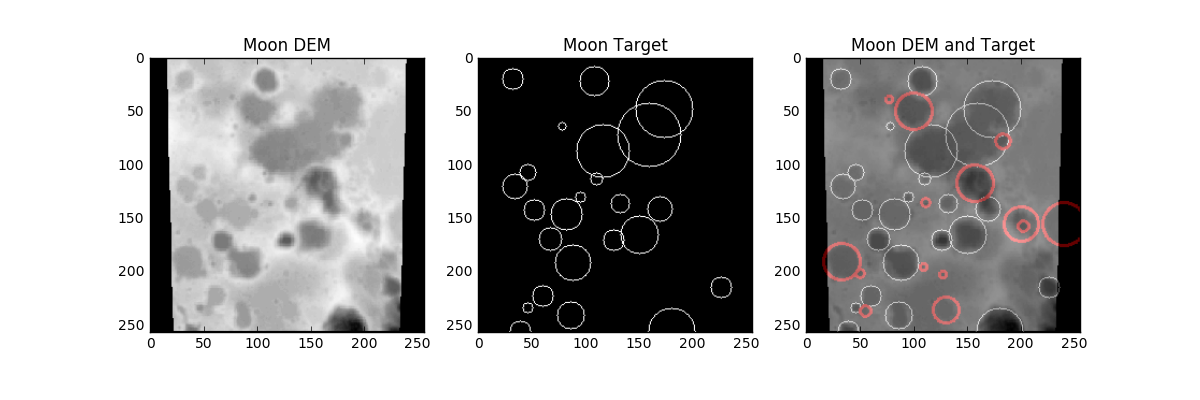}}
\caption{Sample Moon DEM image (left) and target (middle) from our dataset, with the two overlaid (right).
Red circles (right panel) show {features that appear to be craters but are} absent from the \citet{head2010} and \citet{povilaitis2017} datasets, representing apparently missed classifications.}
\label{fig:moondata}
\end{center}
\end{figure}

Each input DEM image is generated by a) randomly cropping a square area of the global map, b) downsampling the cropped image to $256 \times 256\,\mathrm{pixels}$, c) transforming the image to an orthographic projection using the Cartopy Python package \citep{cartopy} in order to minimize image distortion, and d) linearly rescaling image intensity to boost contrast.  The position of the cropped region in a) is randomly selected with a uniform distribution, and its length is randomly selected from a log-uniform distribution with minimum and maximum bounds of $500$ and $6500\,\mathrm{pixels}$ ($59\,\mathrm{km}$ and $770\,\mathrm{km}$), respectively.
The transformation in step c) for our input data often produces non-square images that are padded with zeros; these are the black bars on the sides of the Moon DEM in Figure~\ref{fig:moondata}.\footnote{To check that this padding does not affect our results, we experimented with training our CNN on DEM images with all padding cropped out, and found our performance metrics (Table~\ref{tab:results}) differed by only a few percent. We thus conclude that it has a negligible effect.}

{For each input image, we generate a corresponding ground-truth ``output target'' that is} also $256 \times 256\,\mathrm{pixels}$.  Craters are encoded in the targets as rings with thicknesses of $1\,\mathrm{pixel}$, and with radii and centers derived from craters' physical locations and diameters in the catalog, described below.
All target pixel values are binary, including at ring intersections.
Any craters with diameter $D_\mathrm{pix} < 1\,\mathrm{pix}$ are excluded.
{We experimented with other target formats, including density maps, and binary and non-binary filled circles.  We found, however, that out of all of these formats binary ring targets were best reproduced by the CNN, particularly for situations with many overlapping craters.  While the rings highlight crater edges, the CNN still uses information about the interior of craters to generate the ring masks. We tested this on a limited sample following the method of \\
	 \cite{zeiler,shallue} by perturbing the pixels of input DEMs, and found that we could prevent a crater detection by only increasing the pixels within its interior, without modifying its rim. This demonstrates that the interiors of craters are being used by our network}.

The data used to construct {the targets was} obtained by merging two human-generated crater catalogs.  For $5 - 20\,\rm{km}$ craters we used the global crater dataset assembled by \citet{povilaitis2017} using the LRO Wide Angle Camera (WAC) Global Lunar DEM at $100\,\rm{m/pixel}$ ($303\,\rm{ pixels/degree}$) resolution (GLD100; \citealt{scholten2012}), and for $>20\,\rm{km}$ craters we used the global crater dataset assembled by \citet{head2010} using the LOLA DEM with a resolution of $64\,\rm{ pixels/degree}$ ($472\,\rm{m/pixel}$).  Merging these two datasets in this way was intended by \citet{povilaitis2017}, who explicitly designed their dataset as a continuation of that of \citet{head2010} to smaller sizes.  The methods used to assemble these datasets are described in \citet{head2010} and \citet{povilaitis2017} and are typical for human generated datasets.\footnote{During initial testing we also used the LU78287GT Lunar Crater Catalog \citep{salamuniccar2014}, which was generated by a Hough Transform-based CDA.  We transitioned to solely using human-generated catalogs to prevent the CNN from inadvertently learning the biases of another CDA.}

We split our input image/output target pairs into three separate datasets to be used for training, validating and testing our CNN (see Section~\ref{sec:training} for their uses).  The three datasets are sampled from equal sized and mutually exclusive portions of the Moon, spanning the full range in latitude and $-180^\circ$ to $-60^\circ$, $-60^\circ$ to $60^\circ$ and $60^\circ$ to $180^\circ$ in longitude for the training, validation and test sets, respectively.  Each dataset contains $30000$ DEM images, and the median number of craters per DEM image is $21$.  Because we use a log-uniform sampling of crop lengths, all scales are equally represented in area, but the datasets contain far more DEM images with small crop lengths.  An example input DEM image and target pair is shown in the left and middle panels of Figure~\ref{fig:moondata}.

Our human-generated crater dataset is incomplete and contains many apparent missed identifications. 
\citet{Fasset2012} estimated an incompleteness of $12\%$ for the \citet{head2010} dataset.  The incompleteness of the \citet{povilaitis2017} dataset is unknown at this time, but appears to be comparable or higher.
For the sample DEM image/target pair shown in Figure~\ref{fig:moondata}, we {highlight features that appear to be craters but are} missing from the \citet{head2010} and \citet{povilaitis2017} datasets as red circles in the right panel.
It is unclear at this time why these craters were missed. 
In addition, the right panel of Figure~\ref{fig:moondata} shows how our binary ring targets do not match the rims of non-circular craters.
Together, these hinder the training of our CNN since genuine crater rims will be present in our dataset with no corresponding target rings, potentially confusing the CNN (see Section~\ref{sec:discussion} for a discussion).

\subsection{Convolutional Neural Networks (CNNs)}

In this section we provide a brief, heuristic background to convolutional neural networks (CNNs) before introducing our network architecture.  More in depth descriptions of the theory and mathematics of CNNs can be found in references such as \cite{Goodfellow-et-al-2016} Chapter 9.%, and \cite{hastie2009elements}.

Machine learning algorithms in general can be thought of as universal function approximators, and neural networks (NNs) are a type of machine learning algorithm that uses ``neurons'' (loosely modelled after the human brain) to make such approximations.  A neuron can be represented mathematically by:

\begin{equation}
y = f\left(\sum_j w_j x_j + b\right)
\label{eq:NN}
\end{equation}
where $x_j$ are a set of input values that are linearly combined with a set of weights $w_j$, and added to a bias offset $b$.  This linear combination is {then fed through a (typically non-linear)} ``activation function'' $f$, which returns output $y$.  Depending on the choice of $f(z)$, where $z = \sum_j w_j x_j + b$, a neuron is able to represent a number of simple functions, eg. $f(z) = z$ for a line, or $f(z) = 1 / \left(1 + \exp(-z)\right)$ for a sigmoid function.  Varying the weights $w_j$ and bias $b$ to best approximate a set of known $y$ values given a corresponding set of $x_j$ values is thus equivalent to linear regression when $f(z)$ is linear, and logistic regression when $f(z)$ is sigmoidal.

A neural network contains sets, or layers, of neurons $\left\{y_i\right\}$, where $y_i$ represents the output of the $i$-th neuron in the layer.  Neurons within each layer are independent of one another, but different layers are stacked on top of one another, with the output of one layer serving as the input to the next.\footnote{This is true of standard ``feed-forward'' neural networks.  Other types of networks exist, e.g. recurrent neural networks, but are beyond the scope of this paper.}  Input data is fed into the first layer, while the network's {``prediction'', or predicted output target,} comes from the last.  A network is ``trained'' by tuning the weights of all neurons in all layers so that the network best approximates a set of {known, or ground-truth,} targets given corresponding input.  This tuning is typically done via backpropagation \citep{rumelhart1986}, and goodness of fit is defined through a loss function, e.g. mean squared error between the known targets and network predictions.  Much like adding terms to a Taylor series, increasing the number of layers and/or the number of neurons per layer allows the network to approximate functions of increasing complexity.
However, this added complexity comes at the cost of more tunable parameters and the potential for errors to be amplified from one layer to the next; both make network optimization more difficult.

In computer vision problems, the input is typically pixel intensities from images, while the target is typically either a single class label for the image or another array of pixel intensities. In the latter case (which used in this work), the approximated function is a mapping from one type of image to another. 
Since the input and output are two-dimensional, we may represent the neuron in the i-th row and j-th column of one layer as

\begin{equation}
y_{ij} = f\left(\sum_k\sum_l w_{ij,kl} x_{kl} + b_{ij}\right)
\label{eq:NN2}
\end{equation}
where, for the NN's first layer, $x_{kl}$ represents input image pixel intensities.  Classical NNs, however, place no restrictions on $w_{ij,kl}$, and allow weights of different neurons in a layer to vary independently of one another.  For images, this means any number of pixels in any orientation can be connected, and there is no guarantee the spatial information of the image is preserved from one layer to the next.
 
CNNs primarily differ from traditional NNs in that their neurons are only locally connected to the previous input volume \citep{lecun1989}, and layers are structured to perform a discrete convolution between their inputs and a kernel, or ``filter'', which is represented by the weights.  In the context of Equation \ref{eq:NN2}, this means $w_{ij,kl}$ is zero other than a few adjacent values of $k, l$, and the weights for one neuron are just index-shifted versions of the weights for another.  Convolutional layers hence embed the spatial continuity of images directly into the architecture of the network.  Also, since there are only a few non-zero weights that are shared between all the neurons, only a small number of weights need to be stored in memory and adjusted during training, simplifying network optimization.
The weight-sharing also exploits the property that weights useful to one section of an image might also be useful to another. 

The main components of CNNs relevant to our work are convolutional layers, pooling layers, and merge layers. 
Convolutional layers, the primary component of a CNN, contain neurons that convolve an input with a filter in the manner described above.  {The output of this convolution (i.e. $y_{ij}$ in Equation \ref{eq:NN2}) is called a ``feature map''.}  In practice, a single convolutional layer can contain a large number of filters, each acting on the layer's input {to produce its own feature map}.
%all of the same size but each with different weight values and acting independently from the others on the layer's input.  
We make use of these ``filter banks'' in our CNN architecture.
Pooling layers perform downsampling operations along the spatial dimensions, reducing the dimensionality of the image and {thus the number of learnable weights needed for subsequent layers}.
Merge layers combine {feature maps from} convolutional layers of compatible spatial dimensions, facilitating complex connections within the CNN.  Neither pooling nor merge layers have trainable parameters, and are not represented by Equation \ref{eq:NN2}.

In recent years CNNs have demonstrated impressive performance on image-related tasks, including classifying individual pixels (aka ``segmentation'' in computer vision terminology) \citep{long2015}. 
A particularly successful network for pixel-wise classification (image to image mapping) is the UNET architecture \citep{Ronneberger2015}, which was originally designed for biomedical segmentation problems. 
A novel aspect of the UNET architecture is the use of numerous ``skip connections" which merge deep and shallow layers together, providing both spatial and semantic classification information for future convolutions.

\subsection{CNN Architecture}
\label{sec:cnn}
\begin{figure}
\begin{center}
\centerline{\includegraphics[width=1.0\textwidth]{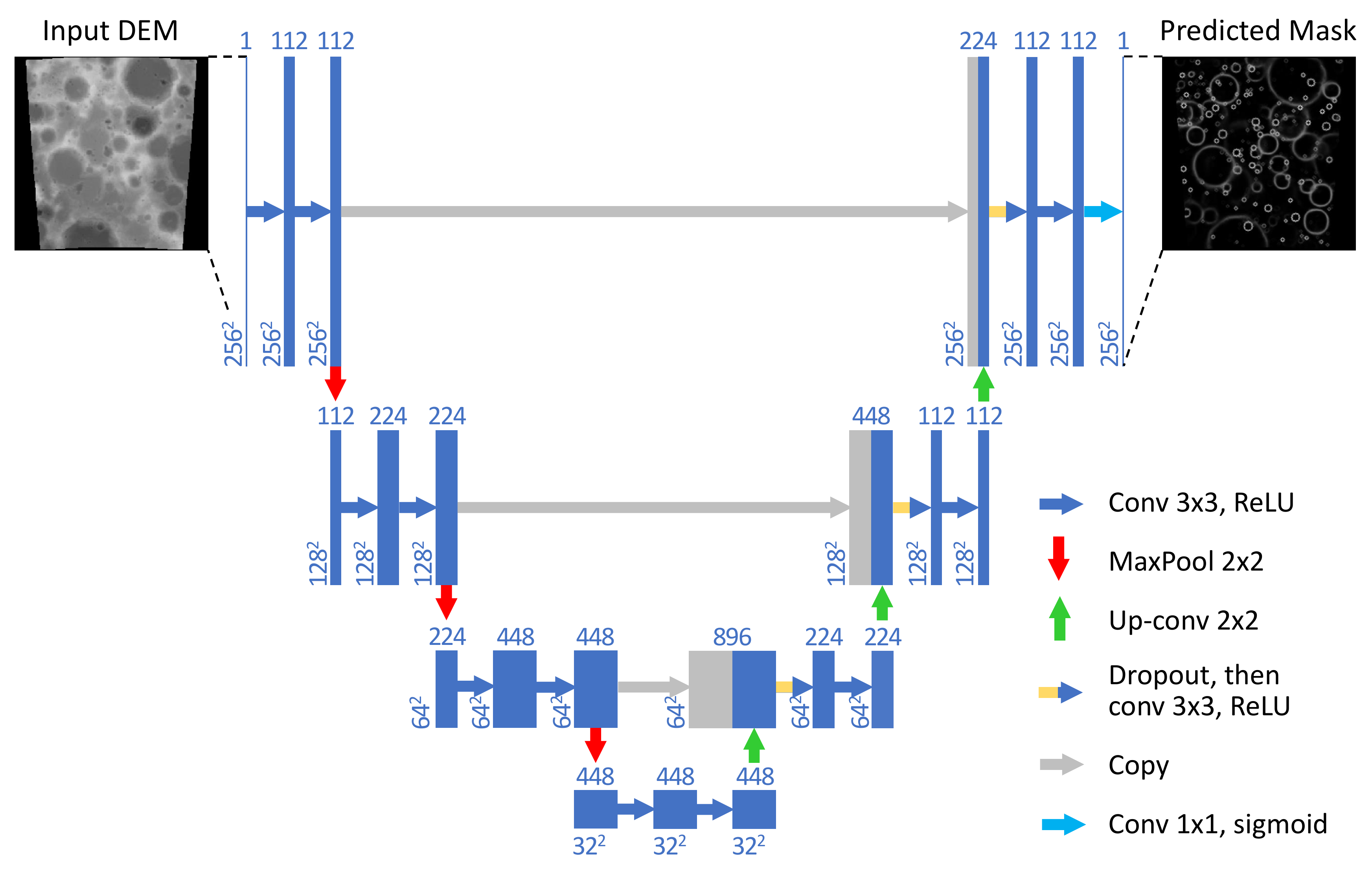}}
\caption{
Convolutional neural network (CNN) architecture, based on UNET \citep{Ronneberger2015}.  Boxes represent cross-sections of {sets of square feature maps.  For each set, its maps' dimensions are indicated on its lower left, and its number of maps are indicated above it.}  Half-grey boxes represent {sets for which half of their maps are copied}.  The leftmost map is a $256 \times 256$ grayscale image sampled from the digital elevation map, and the rightmost the CNN's binary ring mask prediction.  Arrows represent operations, specified by the legend - notably, blue arrows represent convolutions, while gray ones represent copying (skip connections).}
\label{fig:unet}
\end{center}
\end{figure}

In this work we implement a custom version of the UNET architecture \citep{Ronneberger2015}, shown in Figure~\ref{fig:unet}.\footnote{In the early stages of this work, we attempted to count the craters of an image, rather than localize them, using a traditional CNN regressor.  However, that model's resulting accuracy was low, motivating our shift to the UNET-based one.}
This architecture consists of a contracting path (left side) and expansive path (right side), joined through multi-level skip connections (middle).
Lunar DEM images are input to the contracting path and predictions are output from a final layer following the expansive path. 
Unless otherwise stated, all convolutional layers have banks of filters that each apply 3x3 padded convolutions followed by a rectified linear activation unit (ReLU; e.g. \citealt{Goodfellow-et-al-2016} Chapter 6.1), whose functional form is $f(z) = \mathrm{max}\left(0, z\right)$.

The contracting and expansive paths each contain 3 convolutional blocks. 
A block in the contracting path consists of two convolutional layers followed by a max-pooling layer with a 2x2 pool size.  A block in the expansive path consists of a 2x2 upsampling layer, a concatenation with the corresponding block from the contracting path (i.e. a merge layer), a dropout layer \citep{srivastava2014}, and two convolutional layers.  
The connecting path consists of two convolutional layers.   Lastly, the final output layer is a 1x1 convolutional layer with a sigmoid activation and a single filter to output pixel-wise class scores. 
In the contracting path, each convolutional layer in blocks 1, 2 and 3 contain 112, 224 and 448  filters, respectively, while in the expansive path blocks 5, 6 and 7 contain 224, 122 and 122, respectively.
Each convolutional layer in the connecting path contains 448 filters. Our CNN differs from the original UNET by the number of {filters in each convolution layer} (which we selected for the model to fit into GPU memory) and the use of dropout in the expansive path.

\subsection{Crater Extraction}
\label{sec:craterdet}
A $256 \times 256$ DEM image passed though the CNN will output a $256 \times 256$ target with activated pixels corresponding to the locations of the crater rims. 
However, the CNN does not explicitly extract crater position and size from these rims.
Instead, this must be done separately using a custom pipeline that relies heavily on the {\tt{match\_template}} algorithm from {{\tt scikit-image}} \citep{van2014} \footnote{Although template matching is an expensive technique done in a brute force manner we found it far more accurate than others including the Hough transform \citep{Duda1972} and Canny edge detection \citep{canny1986} with enclosed circle fitting.}.
This algorithm iteratively slides generated rings through the targets and calculates a match probability at each ($x$, $y$, $r$) coordinate where ($x$, $y$) is the centroid of the generated ring and $r$ is the radius.

Our custom crater extraction pipeline is as follows.
For each CNN-predicted target we apply a binary threshold $B$ such that pixel intensities greater than $B$ are set to 1 and those otherwise are set to 0. 
We then apply Scikit's {\tt{match\_template}} algorithm over a radius range $r_{\rm{min}}$ to $r_{\rm{max}}$ and classify any ($x$,$y$,$r$) ring with a match probability greater than $P_m$ as a crater.
Two craters $i$ and $j$ that fulfill the following criteria are flagged as duplicates if they satisfy both of the following conditions:

\begin{align}
\begin{split}
((x_i - x_j)^2 + (y_i - y_j)^2) / \mathrm{min}(r_i, r_j)^2 &< D_{x,y} \\
\mathrm{abs}(r_i - r_j)/\mathrm{min}(r_i, r_j) &< D_r 
\label{eq:filt1}
\end{split}
\end{align}
where $D_{x,y}$ and $D_r$ are tunable hyperparameters. 
For duplicates we keep only the crater with the highest match probability $P_m$.
This process is repeated for all CNN-predicted targets.

As is standard practice in machine learning, our hyperparameters are tuned by measuring the performance of various combinations on the validation data and picking the optimal set. After training our CNN (see Section~\ref{sec:training}), we perform a randomly sampled grid search of size 210 on the validation data over the following hyperparameter ranges:

\begin{align*}
\begin{split}
B &= [0.05,0.1,0.15] \\ 
P_m &= [0.3,0.4,0.5,0.6,0.7] \\
D_{x,y} &= [0.6,0.8,1.0,1.2,1.4,1.6,1.8,2.0,2.2] \\
D_r &= [0.2,0.4,0.6,0.8,1.0,1.2]
\end{split}
\end{align*}
We find $B=0.1$, $P_m = 0.5$, $D_{x,y} = 1.8$ and $D_r = 1.0$ yields the optimal $F_1$ score (see Equation~\ref{eq:f1}) of $0.74$.
We set $r_{\rm{min}} = 5$ to minimize errors (see Section~\ref{sec:err}) and set $r_{\rm{max}} = 40$. 

\subsection{Post-Processing}
\label{sec:postproc}
Since our dataset contains DEM images with a log-uniform distribution of magnifications and uniform distribution of locations, a single crater will appear on average in $120 \pm 30$ different DEM images.
This increases the likelihood of detection but also yields many  duplicates across targets which must be filtered.
Therefore, in a final post-processing step we aggregate crater detections across all targets, convert from pixel coordinates to degrees and kilometers, and filter out duplicates. 

Using the known properties of each DEM image, craters are converted from pixel coordinates ($x$, $y$, $r$) to degrees and kilometer coordinates ($\LL$, $L$, $R$):

\begin{align}
\begin{split}
L - L_0 &= \frac{\Delta L}{\Delta H}\left(y - y_0\right) \\
\LL - \LL_0 &= \frac{\Delta L}{\cos\left(\frac{\pi L}{180^\circ}\right)\Delta H}\left(x - x_0\right) \\
R &= r\frac{C_{KD}\Delta L}{\Delta H},
\label{eq:estimate_longlatdiam}
\end{split}
\end{align}
where $\LL$ and $L$ are the crater's longitude and latitude centroid, subscript $0$ values are those for the center of the DEM image, $\Delta L$ and $\Delta H$ are the latitude and pixel extents of the DEM image along its central vertical axis (where $\LL = \LL_0$), excluding any padding, and

\begin{equation}
\label{eq:kmtodeg}
C_{KD} = \frac{180^\circ}{\pi R_\mathrm{Moon}}
\end{equation}
is a kilometer-to-degree conversion factor, where $R_\mathrm{Moon}$ is the radius of the Moon in $\mathrm{km}$.

We then employ a similar filtering strategy as Section~\ref{sec:craterdet}, classifying craters $i$ and $j$ as duplicates if they satisfy both of the following conditions:

\begin{align}
\begin{split}
\frac{\left((\LL_i - \LL_j)^2\cos^2\left(\frac{\pi}{180^\circ}\left\langle L \right\rangle\right) + (L_i - L_j)^2\right)}{C_{KD}^2\mathrm{min}(R_i, R_j)^2} &< D_{\LL,L} \\ 
\frac{\mathrm{abs}(R_i - R_j)}{\mathrm{min}(R_i, R_j)} &< D_R.
\label{eq:filt2}
\end{split}
\end{align}
where $\left\langle L \right\rangle = \frac{1}{2}({L_i + L_j})$. 
$D_R$ and $D_{\LL,L}$ are hyperparameters which, like in Section~\ref{sec:craterdet}, we tune by performing a grid search on the validation dataset after training our CNN and tuning its hyperparameters, this time sampling every combination from:

\begin{align*}
\begin{split}
D_{\LL,L} &= [0.6,1.0,1.4,1.8,2.2,2.6,3.0,3.4,3.8] \\ 
D_R &= [0.2,0.6,1.0,1.4,1.8,2.2,2.6,3.0,3.4,3.8]
\end{split}
\end{align*}
We find $D_{\LL,L} = 2.6$ and $D_R = 1.8$ yields the optimal $F_1$ score (see Equation~\ref{eq:f1}) of $0.67$.

\subsection{Accuracy Metrics}
\label{sec:accuracy}
To train our network we use the pixel-wise binary cross-entropy, $\ell$ \citep{abadi2016a, chollet2015}, a standard loss function used for segmentation problems:
\begin{equation}
\ell_i = x_i - x_iz_i + \textrm{log}(1 + \textrm{exp}(-x_i))
\label{eq:bXE}
\end{equation}
where $z_i$ is the {ground-truth output target value of pixel $i$ and $x_i$ is the CNN-predicted one.}

To optimize the hyperparameters in our crater extraction and post-processing routines (Sections~\ref{sec:craterdet} and \ref{sec:postproc}) we use the precision, $P$, and recall, $R$, to measure accuracy, calculated according to:

\begin{align}
\begin{split}
\label{eq:pr}
P &= \frac{T_p}{T_p + F_p}\\
R &= \frac{T_p}{T_p + F_n}
\end{split}
\end{align}
where $T_p$ are true positives, $F_p$ are false positives and $F_n$ are false negatives. 
There is always a trade-off between precision and recall. 
For example, a machine that only classifies craters when extremely certain will have a high precision but low recall, while a machine that classifiers craters when only moderately certain will have a higher recall but lower precision. 
A common single-parameter metric that balances precision and recall  is the $F_1$ score:
\begin{equation}
\label{eq:f1}
F_1 = 2\frac{PR}{P+R}
\end{equation}

Implicitly encoded into our accuracy metrics is the assumption that our ground-truth datasets of \cite{head2010} and \citet{povilaitis2017} are complete.
However, as mentioned in Section~\ref{sec:data} this is incorrect.
As a result, genuine new craters identified by our CNN will be interpreted as false positives, penalizing these loss functions vs. improving them.
This is an unavoidable consequence of using an incomplete ground truth (see Section~\ref{sec:discussion} for further discussion). 

To measure the accuracy of identified craters (Sections~\ref{sec:craterdet} and \ref{sec:postproc}) we calculate the fractional errors in longitude, $\LL$, latitude, $L$, and radius, $R$, according to:

\begin{align}
\begin{split}
\label{eq:fracerr}
d\LL/R &= \rm{abs}(\LL_P - \LL_G)cos(\pi \left\langle L \right\rangle/180^\circ)/(R_GC_{KD}) \\ 
dL/R &= \rm{abs}(L_P - L_G)/(R_GC_{KD}) \\
dR/R &= \rm{abs}(R_P - R_G)/R_G 
\end{split}
\end{align}
where subscript $P$ corresponds to our CNN-predicted craters, subscript $G$ corresponds to our ground-truth craters and $\left\langle L \right\rangle =\frac{1}{2}(L_P + L_G)$.

Finally, our pipeline discovers thousands of new craters (as will be shown in Section~\ref{sec:Results}). We measure the new crater percentage, $P$, according to:
\begin{equation}
P = N/(N + G)
\label{eq:newcraters}
\end{equation}
where $N$ is the number of CNN-predicted craters without a corresponding match from the ground truth (i.e. they are either genuine new craters or false positives), and $G$ is the number of ground-truth craters.
A ``match" between a CNN-predicted and ground-truth crater is determined via Eq.~\ref{eq:filt1} and Eq.~\ref{eq:filt2} for post-CNN (Section~\ref{sec:craterdet}) and post-processed (Section~\ref{sec:postproc}) craters, respectively.

\subsection{Training}
\label{sec:training}
A recurring theme in machine learning is ``overfitting", which occurs when a model latches onto overly-complex and/or irrelevant features during training. 
Overfit models typically achieve high accuracy on the training set but low accuracy (poor generalization) on new data.
Many algorithms control for overfitting by penalizing overly-complex models, retaining only the essential characteristics from the training data that will generalize to new examples.
This penalization is generally mediated through the model's hyperparameters, which control the model's complexity.
For our CNN, such hyperparameters include weight regularizations for each convolutional layer, the learning rate, dropout layers after each merge layer, filter size, and depth of the network.
These hyperparameters are tuned on a separate validation dataset, forcing the model to achieve high accuracy on two different datasets.

After training the model and tuning the hyperparameters a final evaluation is conducted on the test set, another dataset distinct from both the training and validation datasets.
If the model achieves comparable accuracy on the test set as the training and validation sets, it is likely that minimal overfitting has occurred and the model should generalize well to new examples.
We also address overfitting through a custom image augmentation scheme that randomly flips, rotates and shifts DEM images (and their corresponding targets) before they are used to train the CNN.  This augments the effective dataset size and minimizes the chance of the CNN generating features related to image orientation. 

We tune the hyperparameters of our model by training 60 models with randomly chosen hyperparameters over standard ranges on the training set and selecting the model with the best binary cross-entropy score (Equation~\ref{eq:bXE}) on the validation set. 
Defining an ``epoch" as a single pass through the entire training set and ``batch size" as the number of examples seen per backpropagation gradient update, each model is trained for 4 epochs with a batch size of 8 using the ADAM optimizer \citep{kingma2014}.
The hyperparameters of our best model are weight regularization = $10^{-5}$, learning rate = $10^{-4}$, dropout = 15\%, $3 \times 3$ filter sizes, and a depth of 3. 

\subsection{Errors}
\label{sec:err}

A few sources of error affect the final extracted ($\LL$, $L$, $R$) coordinates of each detected crater.  First, craters can only be detected in pixel increments, and converting from pixels to degrees yields a quantization error, $E_q$, of:

\begin{equation}
E_q = C_\mathrm{offset} \frac{\Delta L}{\Delta H},
\label{eq:quantization_err}
\end{equation}
where $C_\mathrm{offset} \leq 1$ is a constant of order unity representing typical sub-pixel offsets.  Setting $C_\mathrm{offset} = 1$, and considering our largest DEM images ($6500\,\mathrm{pixels}$) where $\Delta L \approx 25^\circ$, we find a maximum quantization error of $E_q \approx 0.1^\circ$, or $\sim3\,\mathrm{km}$.  In principle this error could be reduced by increasing the pixel resolution of each DEM image, though doing so would be memory-intensive.

Second, objects within an orthographic projection become more distorted further from the central longitude and latitude $(\LL_0, L_0)$, which changes the size of smaller craters, and introduces non-circular deformations in larger ones.  Along the central vertical axis, the deviation of the distorted radius from our estimated value using Equation~\ref{eq:estimate_longlatdiam} is

\begin{equation}
F_{rd} = \frac{R_\mathrm{distorted}}{R} = \frac{1}{\cos\left(\frac{\pi}{180^\circ}(L - L_0)\right)}.
\label{eq:raddistortion_frac_err}
\end{equation}
For our largest DEM images, $L - L_0 \approx 12^\circ$, $F_{rd} \approx 1.02$, so deviations are at most $2\%$ of a crater's radius.

Third, the longitude and latitude estimates in Equation~\ref{eq:estimate_longlatdiam} neglect higher order terms (including the distortion described above) and cross-terms containing both $\LL$ and $L$.  To quantify this effect we passed the crater pixel positions from the ground-truth test dataset through Equation~\ref{eq:estimate_longlatdiam} to obtain $\LL_C$ and $L_C$.  We then subtracted the ground truth's longitude and latitude values from $\LL_C$ and $L_C$, respectively, and normalized by the longitude/latitude extent of our DEM images.  We found median relative offsets of $0.13$\% and $0.28$\% in longitude and latitude, but, for the largest DEM images, maximum relative offsets can reach $1.0$\% in longitude and $1.9$\% in latitude.  For a $6500\,\mathrm{pixel}$ DEM image, this translates to $0.25^\circ$ in longitude and $0.5^\circ$ in latitude.
%, which, while too large for producing an accurate crater catalog, remain smaller than the duplicate threshold in Equation~\ref{eq:filt2}.  
We also calculated the fractional error using Equation~\ref{eq:fracerr}, replacing $\LL_P$ and $L_P$ with $\LL_C$ and $L_C$, and find median fractional errors of $3\%$ in longitude and $5\%$ in latitude.%, less than half their values in Table~\ref{tab:results}.

To help offset these errors we impose a minimum search radius, $r_{\rm{min}} = 5$, for our crater extraction pipeline (Section~\ref{sec:craterdet}).
This prevents quantization and projection errors from ever being a significant fraction of the crater's radius. 
This comes at a cost of not being able to probe the smallest craters in each DEM image, yielding fewer new crater detections than we otherwise would obtain. 

\section{Results}
\label{sec:Results}
\subsection{Crater Identification on the Moon}
We apply our trained CNN and optimized crater identification pipeline on the test set and list our various accuracy metrics in Table~\ref{tab:results} for the validation and test sets. ``Post-CNN" statistics were generated on an image-by-image basis after Section~\ref{sec:craterdet} of the pipeline with an averaged mean and standard deviation taken across all predicted targets. ``Post-processed" statistics were generated after Section~\ref{sec:postproc} of the pipeline, and hence represent our final crater distribution after combining extracted craters from each target into one distribution and removing duplicates. Together, these statistics convey how our pipeline is performing at various stages.

\newcolumntype{P}[1]{>{\centering\arraybackslash}p{#1}}
\begin{table}
\small
\begin{center}
\begin{tabular}{ |P{3.9cm}|P{2.1cm}|P{2.1cm}|P{1.9cm}|P{1.9cm}| } 
\hline 
Accuracy Metric & Post-CNN (Validation) & Post-Processed (Validation) & Post-CNN (Test) & Post-Processed (Test) \\ \hline
 Recall & $56\% \pm 20\%$ & $92\%$ & $57\% \pm 20\%$ & $92\%$\\
 Recall ($r<15$ pixels) & $83\% \pm 16\%$ & -- & $83\% \pm 13\%$ & -- \\
 Precision & $81\% \pm 16\%$ & $53\%$ & $80\% \pm 15\%$ & $56\%$ \\
 New Crater Percentage & $12\% \pm 11\%$ & $45\%$ & $14\% \pm 13\%$ & $42\%$ \\
 False Positive Rate & -- & -- & -- & $11\% \pm 7\%$\\
 Frac. longitude error & $10\%^{+2\%}_{-2\%}$ & $13\%^{+10\%}_{-7\%}$ & $10\%^{+2\%}_{-2\%}$ & $11\%^{+9\%}_{-6\%}$ \\
 Frac. latitude error & $10\%^{+3\%}_{-2\%}$ & $10\%^{+8\%}_{-5\%}$ & $10\%^{+2\%}_{-2\%}$ & $9\%^{+7\%}_{-5\%}$ \\
 Frac. radius error & $8\%^{+2\%}_{-2\%}$ & $6\%^{+5\%}_{-3\%}$ & $8\%^{+1\%}_{-1\%}$ & $7\%^{+5\%}_{-4\%}$  \\
% Binary Cross-Entropy Loss & $0.096$ & --  \\
 \hline
\end{tabular}
\caption{Accuracy metrics on the validation and test sets. ``Post-CNN" statistics were generated after Section~\ref{sec:craterdet} of the pipeline with a mean and standard deviation taken across targets, while ``post-processed" statistics were generated after Section~\ref{sec:postproc} of the pipeline, after combining extracted craters into a single global distribution. 
Precision and recall are calculated according to Eq.~\ref{eq:pr}, new crater percentage according to Eq.~\ref{eq:newcraters}, fractional longitude, latitude and radius errors according to Eq.~\ref{eq:fracerr} with a median and interquartile range (IQR) taken across all detections. The false positive rate of new craters is estimated by four different scientists classifying 361 new craters and averaging the results. We note that precision drops at the post-processed stage because many new craters (absent in the ground truth) are identified.
}
\end{center}
\label{tab:results}
\end{table}
The similarity between our validation and test set statistics in Table~\ref{tab:results} implies that little to no overfitting has occurred. 
Our post-processed test recall is $92\%$, recovering almost all craters from the test set. 
By comparison, our post-CNN test recall is lower at $57\% \pm 20\%$, meaning that (on average) our CNN detects only half of the craters per target.
The drastic difference between post-processed and post-CNN recalls demonstrates the effectiveness of aggregating crater detections across multiple images and scales.
A major reason for our low post-CNN recall is our CNN does not reliably detect craters with radii greater than $\sim 15$ pixels (see Section~\ref{sec:discussion} for a discussion). 
Restricting to craters with a pixel radius $r$ less than $15 \ \rm{pixels}$, our post-CNN test recall improves to $83\% \pm 16\%$. 
\citet{wetzler2005} estimated a human recall of $75\%$ when re-classifying crater images, making our post-CNN recall consistent with human performance for $r<15$ pixels.

$42\%$ of post-processed test craters are new, almost doubling our catalog, with $15\%$ of them having diameters under $5\,\rm{km}$ (i.e. below the limits of our ground-truth catalogs).
Our estimated false positive rate of these new post-processed craters is $11\% \pm 7\%$, which was estimated by four scientists from our research group each classifying the same 361 new craters re-projected onto their original DEM images and averaging the results. 
This procedure allowed the scientists to classify the new craters under the same conditions as our identification pipeline.
These manually classified craters along with their corresponding Moon DEMs, ground-truth targets and CNN predictions are publicly available at \url{https://doi.org/10.5281/zenodo.1133969}.
Although individual false positive estimates differed between scientists, this is in line with previous research \citep[e.g.][]{robbins2014} that large disagreements in human crater classification is common. 
For post-CNN, $14 \pm 13\%$ of test craters per DEM image are new.
As a result of these new crater detections, our post-CNN and post-processed precisions are low since new craters are interpreted by the precision metric exclusively as false positives (see Section~\ref{sec:discussion} for a discussion).

\begin{figure}
\begin{center}
\centerline{\includegraphics[width=1.05\textwidth]{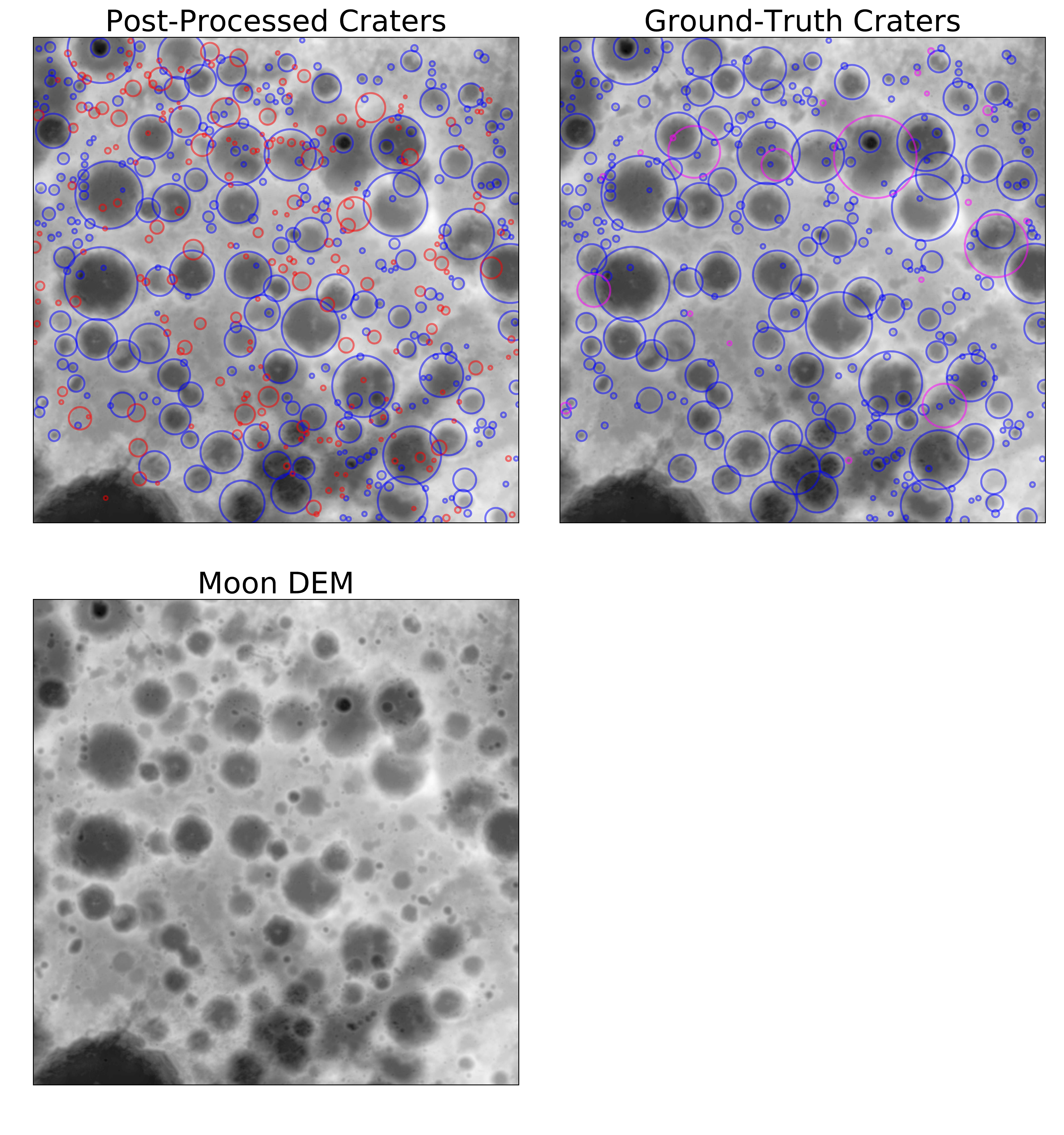}}
\caption{Sample patch of the Moon from the test set (lower left), with post-processed (top left) and ground-truth (top right) craters overlaid. Blue circles represent post-processed craters that were successfully matched to the ground truth (and vice versa), red circles represent new crater detections from our pipeline (without a corresponding ground-truth match), and purple circles represent ground-truth craters missed by our pipeline.
}
\label{fig:cratercomp}
\end{center}
\end{figure}

Figure~\ref{fig:cratercomp} compares our post-processed craters (top left) to the ground truth (top right) for a large swath of the Moon (bottom left) from the test set. 
Blue circles represent post-processed craters that were successfully  matched to the ground truth (and vice versa), red circles represent new crater detections from our pipeline (without a corresponding ground-truth match), and purple circles represent ground-truth craters missed by our pipeline. 
As can be visually seen, our pipeline recovers many more craters than the ground truth, with overall few false positives and duplicates. 
Our median post-processed and post-CNN fractional errors in longitude, latitude and radius are $11\%$ or less, representing overall good agreement with the ground truth despite the sources of error mentioned in Section~\ref{sec:err}.

\begin{figure}
\begin{center}
\includegraphics[scale=0.4]{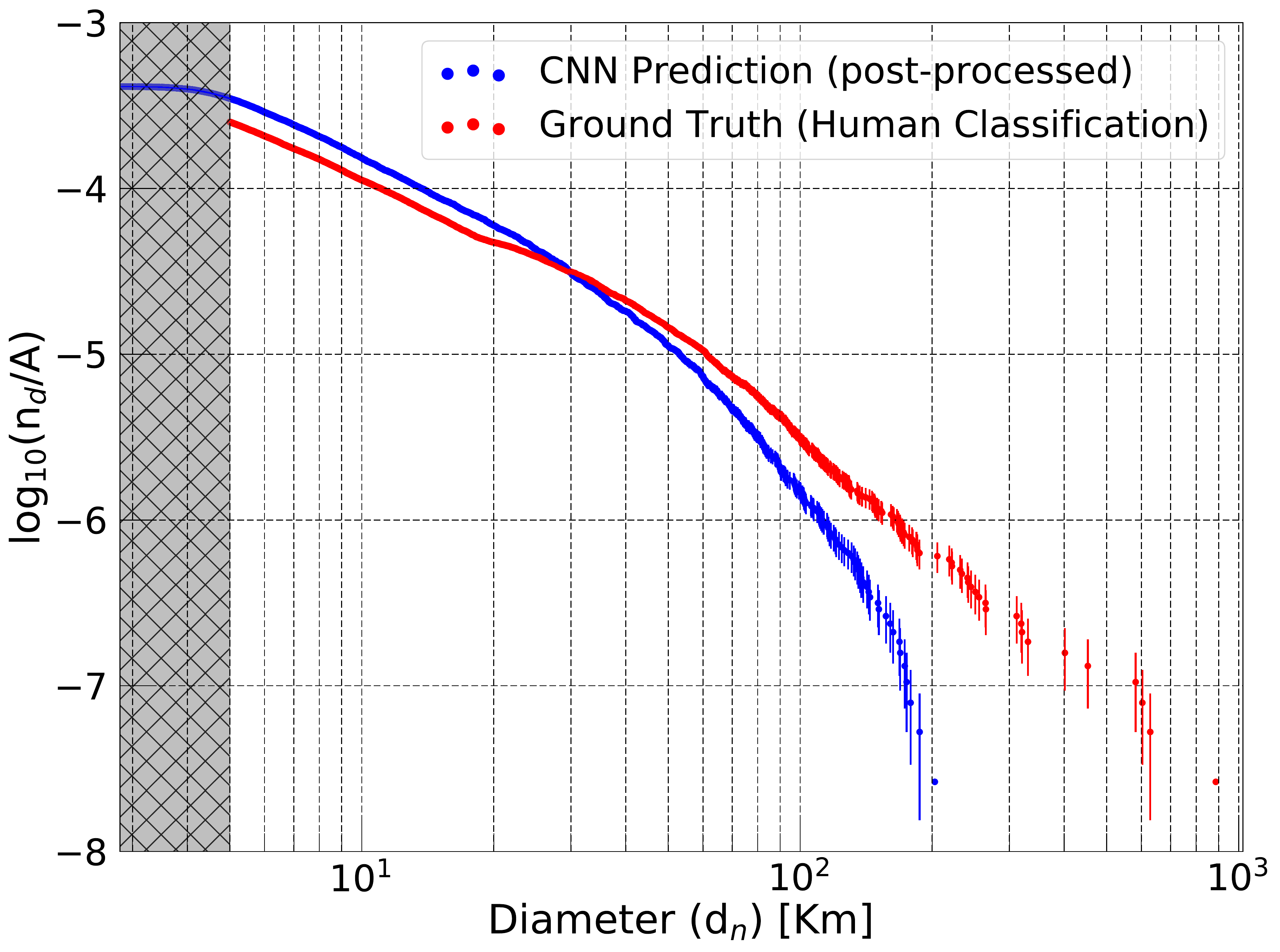}
\caption{{Lunar crater size-frequency distributions represented as CSFD plots. Red is the human-generated test set (the ground truth), compared to what our CNN predicts post-processing (blue). Our CNN recover the same slope as the GT for craters smaller than 20~km while detecting significantly more craters in this range. The shaded region inside 5 km however should not be physically interpreted due to data incompleteness.}}
\label{fig:main_plot}
\end{center}
\end{figure}

\subsection{Lunar Crater Size Distribution}
\label{sec:csfd}

{In Figure~\ref{fig:main_plot} we show the cumulative size-frequency distribution (CSFD) of the craters recovered by our CNN as compared with the CSFD of craters in the human counted set.  This is constructed following the recommendations in NASA Technical Memorandum (79730, 1978).  In this case the surface area, $A$, used to normalise the number of craters is the area of the Moon between latitudes of 60$^{\circ}$ North and South. We note that another widely used method to visualize crater distributions is R plots (as for example used in \cite{head2010}), however we refrain from using these due to their significant dependency on data binning that can be completely avoided using CSFDs, where each crater is its own bin \citep{weaver}. }

{As we can see, between roughly 5 and 20~km diameters the CNN derived CSFD is systematically higher than the human derived CSFD, but runs essentially parallel to it.  This indicates that while the CNN finds a substantially larger number of craters than the human crater counters, the craters newly identified by the CNN follow the same size distribution as the human identified craters, which reassures us that they are likely to be real.  At 20~km however there is a slight kink and upturn in the human dataset whereas the CNN prediction continues smoothly.  The CNN prediction then curves downward gradually before cutting off more sharply at diameters of around 200~km, resulting in increasing divergence between the CNN prediction and the human classifications for these larger craters.}

{There are several factors that might contribute to the divergence between the CSFDs of the CNN prediction and the human classified dataset at large crater sizes.  Firstly, as we describe in detail in Section~\ref{sec:discussion}, our CNN can struggle with identifying craters that have a radius of more than around 15~pixels in the input images.  While this is somewhat ameliorated by sampling images at many different magnification levels (such that an $r>15$~pix crater in one image will have $r<15$~pix in another) it can still slightly hinder the detection of craters with diameters larger than around 50~km.  This is likely responsible for some of the downward curvature of the CNN CSFD at larger sizes and is probably the cause of the cut-off at $\sim$200~km.}

{The location of the kink in the human classified CSFD at which the mismatch between the CNN prediction and the human classification begins is also notable.  This kink appears at the boundary between the two component datasets of our human classified ground truth, that of \citet{povilaitis2017} and \citet{head2010}.  It is possible that there is a systematic difference between the crater classifications of the two groups.  Indeed some level of difference would be unsurprising given the findings of \citet{greeley1970}, \citet{kirchoff2011} and \citet{robbins2014}.  The systematic offset between the CSFDs of our CNN predictions and the human classifications in the range where the two are parallel already indicates that the human datasets are not complete, as was also found by \citet{Fasset2012} for the \citet{head2010} dataset.  Since the \citet{povilaitis2017} dataset contains around four times as many craters it is inevitable that if there are any systematic differences in the human classification that lead to size-dependent completeness effects, the CNN will tend to follow \citet{povilaitis2017}.}

{Another consideration is that 20~km is roughly the diameter at which craters on the Moon transition from simple to complex \citep[e.g.][]{pike1980, grieve,stoffler}.}  {Above this size crater morphologies change, with crater floors becoming flatter and the appearance of features like central peaks.  It is conceivable that this morphological transition could hinder the detection of larger craters by the CNN.  In particular, since smaller, simple craters dominate the human dataset, if the morphological differences make it difficult to simultaneously optimise to detect both simple and complex craters it is inevitable that the CNN will favour simple craters, just as it would favour \citet{povilaitis2017} over \citet{head2010} in the case of differences in their counting methods.  In Section~\ref{sec:discussion} we discuss how one could attempt to disentangle these possible effects.}

{Below around 5~km diameter the CNN prediction begins to roll over.  This is due to incompleteness in the sampling of the lunar surface at the smallest scales/largest image magnifications.  Despite incompleteness in the sampling and the lack of craters in the training set at these small sizes the CNN still finds many craters $<$5~km diameter.}

\subsection{Transfer Learning on Mercury}
Domain shift is a common problem in machine learning that arises when a model is used to predict on data with different properties than its training set, and typically results in decreased performance.
We briefly evaluate the sensitivity to domain shift for our network by taking our Moon-trained CNN and transfer learning to Mercury.
Mercury has different properties than the Moon, including a different gravitational acceleration, surface composition, terrain, and impact history.
In addition, we also use the Mercury MESSENGER Global DEM with a resolution of $64\,\mathrm{pixel/degree}$, or $665\,\mathrm{m/pixel}$ (\citealt{becker2016}; available at \citealt{messenger2016}), which has different image properties than our Moon DEM. 
All these affect the distribution and appearance of impact craters. 

\begin{figure}
\begin{center}
\centerline{\includegraphics[width=1.4\textwidth]{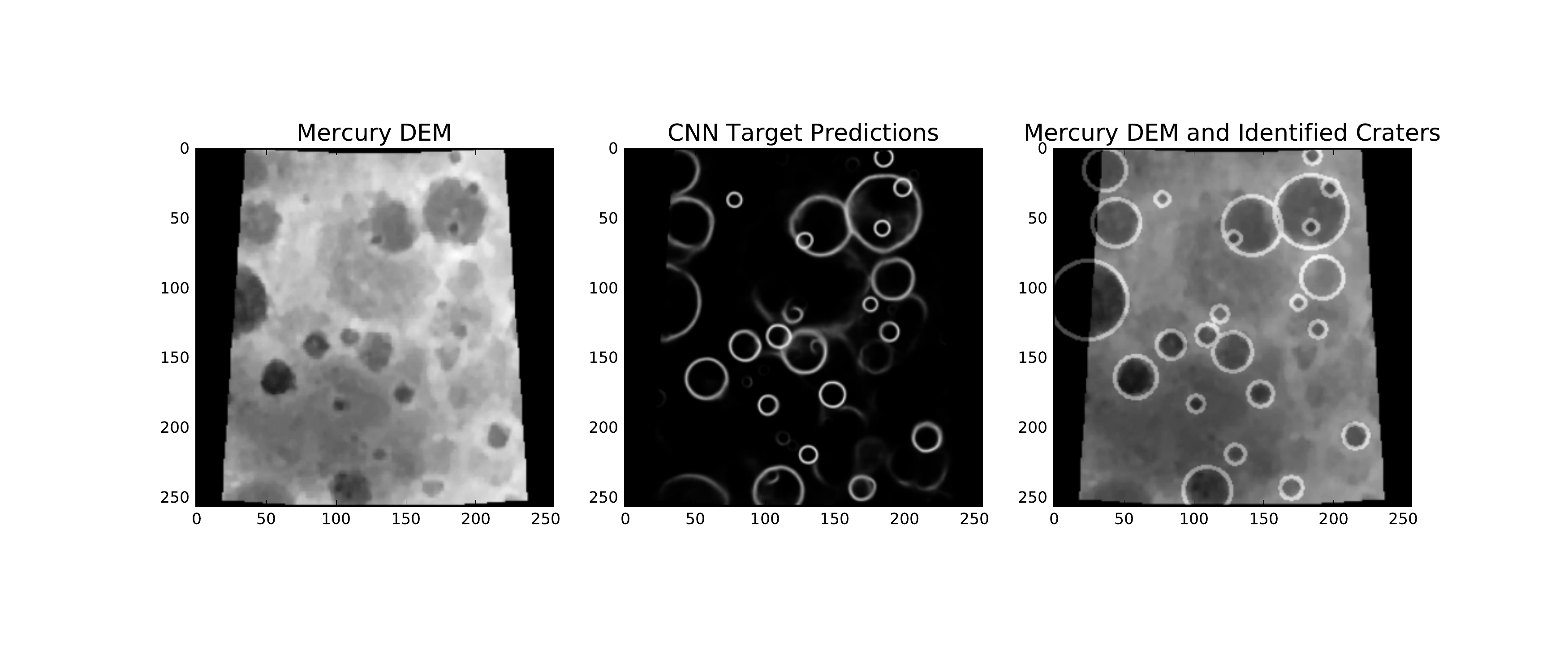}}
\caption{A sample Mercury DEM (left), CNN target predictions (middle), and post-processed identified craters overlaid on the original DEM image (right).}
\label{fig:mercury}
\end{center}
\end{figure}

To evaluate our CNN on Mercury, we prepared DEM images from the Mercury MESSENGER Global DEM in a similar manner as described in Section~\ref{sec:data} (except that we do not use a corresponding human-generated crater catalog).  We then passed these DEM images through our Moon-trained CNN with no alterations to the architecture or weights.  Figure~\ref{fig:mercury} shows a sample input DEM image of Mercury (left), CNN target predictions (middle), and post-processed identified craters overlaid on top of the original DEM image (right).
Comparing the left panels of Figure~\ref{fig:moondata} and Figure~\ref{fig:mercury}, some differences are visible between Moon and Mercury craters, yet our CNN is able to detect both types.
{In addition, as shown in Figure~\ref{fig:mercury}, the CNN correctly identified a significant number of the crater-like features on Mercury’s surface. Moreover, manual inspection on that patch of Mercury shows that almost all of the inferences made by the CNN do seem to be real craters.} This demonstrates  its efficiency in distinguishing craters from other terrain features. Simple edge detection techniques would not make such a distinction. {While humans are often very good at transfer learning, and thus it may seem simple from a human perspective, transfer learning has commonly been challenging in machine learning.  That our model seems to transfer well is thus greatly encouraging.}
While this demonstrates successful generalization, we leave a thorough analysis of transfer learning to future work. 

\section{Discussion} 
\label{sec:discussion}
There are many reasons to believe our CNN has learned the complex features that define a crater.
First, despite the Moon's large structure variations across its surface our CNN was able to recover $92\%$ of craters on a face previously unseen by our CNN.
Second, the similarity in accuracy metrics between the validation and test sets in Table~\ref{tab:results} implies that minimal overfitting has occurred, and the CNN has indeed learned useful, generalizable features. 
Third, $42\%$ of extracted Lunar craters are new, and from human validation of a subset most appear to be genuine matches. 
Fourth, our Moon-trained CNN successfully detected craters on Mercury, a surface completely distinct from any specific region on the Moon.
Finally, while simple edge detection techniques would activate non-crater features like mountains, ridges, etc., our CNN almost  exclusively activates crater rims (e.g. see middle panel of Figure~\ref{fig:mercury}).

As mentioned in Section~\ref{sec:data} and shown in Figure~\ref{fig:moondata}, our training data is incomplete, containing many missed craters as well as target rings that differ from true crater rims. 
Despite these shortcomings our CNN was still able to understand our training objective and correlate the binary ring targets with the true rims of the craters.
Proof of this can be seen in the middle panel of Figure~\ref{fig:mercury}, where some CNN-predictions are non-circular and better match the true crater rims than a circular ring could\footnote{To be clear, the $256 \times 256 \ \rm{pixel}$ CNN target predictions can produce non-circular ring boundaries, as shown in the middle panel of Figure~\ref{fig:mercury}. However, extracted post-processed craters (Section~\ref{sec:craterdet} and Section~\ref{sec:postproc}) do not retain non-circularity information, as shown in the right panel of Figure~\ref{fig:mercury}.}.
Together, these highlight the robustness and flexibility of deep learning solutions.

A fundamental difficulty when using an incomplete dataset is tuning the hyperparameters. 
Under this regime genuine new crater detections are interpreted as false positives, penalizing the precision metric and artificially lowering the $F_1$ score (which we are trying to maximize). 
Since thousands of new craters were detected, the $F_1$ score, which favors hyperparameters that yield the fewest new crater detections whilst still maintaining a high recall, is reduced. 
As a result, our tuning procedure yields hyperparameters that are likely conservative compared to if we had focused on finding new craters or had a more complete ground truth. 
The same principle applies when using the binary cross-entropy loss to train our CNN, yielding a final model that likely generates more conservative crater predictions. 
Followup work that uses a more complete ground-truth crater distribution would presumably yield improved results. 

Our CNN robustly detects craters from each DEM image with radii below 15 pixels, but tends to miss larger craters. 
An example of this can be seen in Figure~\ref{fig:mercury}, where a large crater with coordinates ($x=130, y=78, r=42 \ \rm{pix}$) is only partially recognized by the CNN and missed by our crater extraction pipeline.
We believe that this largely stems from the scale that is imposed when using small 3x3 filters in our convolutional chain, yielding a receptive field that is too small for large craters.
Larger filter sizes were attempted, but this dramatically increases both the number of trainable parameters and network size, making model-optimization more difficult.
Dilated Convolutions \citep{yu2015}, larger convolution strides, and/or deeper networks are possible avenues for improvement. 
However, increasing the receptive field likely accomplishes the same effect as reducing the magnification of an image, which we have already shown to be a successful remedy, achieving a post-processed recall of $92\%$ (see Table~\ref{tab:results}).

{In addition, as we noted in Section~\ref{sec:csfd}, there are potential external effects that may be feeding into the apparent size dependence of the ability of the CNN to recover larger craters.  One possibility is that there is a systematic difference in the two human counted datasets that we have stitched together to form our ground truth.  Another is that the physical transition between simple and complex craters makes it difficult for the CNN to adapt to detect both with equal efficiency.  In both cases the CNN would tend to adapt itself to better identify craters in the larger population, which is the \citet{povilaitis2017} dataset that consists of craters below the $\sim$20~km simple to complex transition.  To disentangle these two possible effects requires a uniformly generated human dataset of both simple and complex craters.  We would then be able to train the CNN on DEMs of solely simple craters, solely complex craters, or both and then test it on DEMs of all physical scales/crater morphologies.  This requires a substantial new human counting effort and so we leave it for future work.}

Our estimated post-processed false positive rate of new craters is $11\% \pm 7\%$, which, although generally low, is likely too high for our catalog to be used to produce high-precision crater catalogs.  
Our primary false positives are a) ambiguous circular looking {depressions} that may or may not be true craters (further analysis required), and b) overlapping craters that activate enough pixels in the region to breach the match probability threshold $P_m$, creating a ``ghost" detection in our crater extraction pipeline (Section~\ref{sec:craterdet}).  In addition, Table~\ref{tab:results} shows that roughly $25\%$ of post-processed craters have coordinates that differ from the ground truth by $20\%$ or more, and examples of this can be seen in Figure~\ref{fig:cratercomp}.  This higher-error tail is not present in the post-CNN errors, so they arise from the post-processing methods, whose sources of error are detailed in Section~\ref{sec:err}.  These issues indicate the need for further refinements to our overall crater identification pipeline in order to produce precision crater catalogs, which we save for future work.

\section{Conclusions and Future Work}
In this work we have demonstrated the successful performance of a convolutional neural network (CNN) in recognizing Lunar craters from digital elevation map images. 
In particular, we recovered $92\%$ of craters from our test set, and almost doubled the number of total crater identifications.
Furthermore, we have shown that our Moon-trained CNN can accurately detect craters on the substantially different DEM images of Mercury. 
This implies that the CNN has learned to robustly detect \textit{craters}, and not features particular to the planetary surface on which it was trained.

Two primary advantages of a deep learning solution over human crater identification are consistency and speed.
A CNN will classify an image identically each time, but the same is not true for humans \citep{wetzler2005}.
In addition, different humans will use slightly different criteria, which adds to the error budget \citep{robbins2014}.
Once trained, our CNN greatly increases the speed of crater identification, taking minutes to generate predictions for tens of thousands of Lunar DEMs and a few hours to extract a post-processed crater distribution from those DEMs.
This is of course all done passively, freeing the scientist to do other tasks. 
Our CNN could also be used to assist human experts, generating initial suggestions for the human expert to verify.

DEMs are available for many other Solar System bodies, including Mercury \citep{becker2016}, Venus \citep{magellan1997}, Mars \citep[][]{fergason2017}, Vesta \citep{preusker2014} and Ceres \citep{preusker2017}. 
It will be interesting to study to what extent our CNN can transfer-learn to other Solar System bodies with a DEM, possibly facilitating a systematic, consistent, and reproducible crater comparison across Solar System bodies. 
While we have successfully shown transfer-learning from our Moon-trained CNN to Mercury, a detailed analysis for Mercury has been left to future work.

Our current work detected craters down to roughly 3 km diameter, but since our CNN accepts images of arbitrary magnification we can transfer-learn to kilometer and sub-kilometer scales on the Moon. We anticipate that the uncharted territory of systematic small-size crater identification will provide important new information about the size distribution of Lunar impactors and the formation history of the Moon. 

\section*{Acknowledgments}
This work was supported by the Natural Sciences and Engineering Research Council of Canada. Computations were performed on the P8 supercomputer at the SciNet HPC Consortium. SciNet is funded by: the Canada Foundation for Innovation under the auspices of Compute Canada; the Government of Ontario; Ontario Research Fund - Research Excellence; and the University of Toronto. The authors are grateful to Fran\c{c}ois Chollet for building Keras.  The authors thank R. Povilaitis and the LROC team for providing the 5-20~km dataset and useful discussions regarding the input data, R. Malhotra and R. Strom for providing helpful comments on a draft of the manuscript, and N. Hammond for useful discussions on craters distributions.  Finally, the authors thank David Minton and an anonymous referee for their constructive comments during the review process that significantly improved the quality of this manuscript.

%% The Appendices part is started with the command \appendix;
%% appendix sections are then done as normal sections

%\end{linenumbers}
%% \section{}
%% \label{}

%% If you have bibdatabase file and want bibtex to generate the
%% bibitems, please use
%%
  \bibliographystyle{elsarticle-harv} 
  \bibliography{craters.bib}

%% else use the following coding to input the bibitems directly in the
%% TeX file.

\end{document}